# ACRR: Ad-hoc On-Demand Distance Vector Routing with Controlled Route Requests

**Jayesh Kataria***

Mumbai University, India, jayeshkataria@gmail.com

**P.S. Dhekne**

BARC, Mumbai, India, dhekne@barc.gov.in

**Sugata Sanyal**

TIFR, Mumbai, India, sanyal@tifr.res.in

**Corresponding Author*

**Abstract**

Reactive routing protocols like Ad-hoc On-Demand Distance Vector Routing (AODV) and Dynamic Source Routing in Ad-Hoc Wireless Networks (DSR) which are used in Mobile and Ad-hoc Networks (MANETs) work by flooding the network with control packets. There is generally a limit on the number of these packets that can be generated or forwarded. But a malicious node can disregard this limit and flood the network with fake control packets. These packets hog the limited bandwidth and processing power of genuine nodes in the network while being forwarded. Due to this, genuine route requests suffer and many routes either do not get a chance to materialize or they end up being longer than otherwise. In this paper we propose a non cryptographic solution to the above problem and prove its efficiency by means of simulation.

**Keywords**

Ad-hoc network, wireless, Routing, Manets, Flood Control, AODV

## 1. INTRODUCTION

Route formation in MANETs has been studied for many years and several protocols like DSR [1], AODV [2], and DSDV [3] have been proposed and optimized to make the task efficient. However, ensuring security in such networks is a big challenge because of the distributed nature of these networks and the assumption of mutual trust among participants. A broad overview of the security issues involved in Ad-hoc networks has been provided in [4]. Some problems are handled by approaches like Ariadne [5], S-AODV [6], SEAD [7], etc which use cryptographic techniques. In this paper we focus on solving the problem of flooding of fabricated packets in the network by using non-cryptographic techniques.

Flooding of fake route requests in the network by malicious nodes can lead to a lot of wastage of resources. With the help of simulations using the AODV protocol, we show that flooding of route requests causes significant throughput degradation and disruption of route formation. In [8], the authors have proposed a simple scheme to limit the request packets forwarded by a neighbor. In this paper we reiterate the solution we proposed in [9]. Additional simulation results further strengthen the efficiency of our solution.

The following section explains the problem we try and solve. Section 3 briefly explains the solution proposed in [8] while in Section 4 we recap our proposed scheme. Extended simulation results, proving our solution are presented in Section 5. In Section 6 we prove the applicability of our results to other reactive routing protocols using logical reasoning. Section 7 narrates the future work possible and concluding remarks.

## 2. PROBLEMS DUE TO ROUTE REQUEST FLOODING

AODV facilitates route formation using control messages RREQ (route request) and RREP (route reply). Each time a data packet is to be delivered by a node, the node checks whether it has a route to the destination. If it does not have a route, then a RREQ is broadcast to the neighbors. The neighbors rebroadcast the packet if they do not know the destination. In order to control the number of RREQ packets generated by a node, the AODV protocol specifies a parameter RREQ_RATELIMIT. However, a malicious node can choose not to observe this limit and flood the network with a large number of fabricated RREQ packets which keep getting forwarded. This process continues, allowing the fake RREQ packets to propagate through the network. As the number of fabricated RREQ packets sent by the malicious nodes increases, the other nodes in the network use up their RREQ_RATELIMIT in forwarding these while other genuine RREQ packets are dropped.

The phenomenon explained above has various side effects ranging from inefficient routing to complete blocking of route formation. The route forming process is disrupted severely in the vicinity of the malicious node and the impact decays slowly as we move away from the malicious node. The fabricated RREQ packets to be processed at a given node outnumber the genuine RREQ packets, which increases the probability of the non-malicious RREQ not being forwarded because of the RREQ_RATELIMIT constraint. The non-malicious nodes do not form routes at all or end up forming longer routes as they try to avoid the high contention region near the malicious node.

Thus the effects of flooding can be summarized as follows:
- Wastage of memory while maintaining routing table entries for malicious requests
- Wastage of battery power
- Denial of service to genuine nodes
- Creation of longer routes where short ones could have been possible leading to

reduced throughput (throughput decreases with increase in hop count)
- Consumption of limited processing power

It thus becomes very important to avoid or at least contain such flooding attacks so as to allow genuine routes to be formed in the network and also to conserve the limited resources available to the mobile nodes.

## 3. INITIAL NAIVE SOLUTION

In this section we summarize the initial naïve approach proposed in [8]. It is a simple and intuitive algorithm to detect and isolate malicious nodes. In this approach, malicious nodes are detected and isolated on the basis of the following three constants.

1. ***RREQ_ACCEPT_LIMIT (RAL)***: It is the number of RREQ packets that can be accepted and processed per unit time by a node from each neighbor. This value tries to be fair to all neighbors by accepting a few RREQ packets from all rather than many from only one neighbor.

2. ***RREQ_BLACKLIST_LIMIT (RBL)***: It is the threshold value that aids each node in determining if a neighbor is acting malicious. If the RREQ packets sent by the neighbor per unit time exceed this value, the neighbor is deemed malicious and blacklisted.

3. ***BLACKLIST_TIMEOUT (BT)***: It is the time for which a malicious node gets blacklisted. It increases exponentially as a particular malicious node gets blacklisted frequently.

The effect of this approach is that if some node starts behaving malicious and tries to flood the network by sending out fake RREQ packets, its activities are detected by its neighbor who is receiving and keeping a track of these and (possibly) other RREQ packets. Once this malicious node crosses *RBL*, its neighbor blacklists it for *BT*. Once this happens all RREQ packets sent by this node are rejected. This enables the neighbor node to process genuine RREQ packets sent by other nodes.

The process of blacklisting the malicious node as explained above is performed by all the neighbors surrounding it. Because all neighbors blacklist the malicious node, it gets isolated and its fake RREQ packets are no longer forwarded along further hops and cannot flood the network. Hence nodes can process and forward genuine RREQ packets leading to the formation of valid and shorter routes.

The flaw in this scheme is the simplistic assumption of *RAL* and *RBL*. These have been predefined and cannot be dynamically varied based on the network topology and other available resources like memory and battery power. Consequently these constants do not always suit all possible scenarios. This can lead to problems like blacklisting of normal nodes or dropping more RREQ packets than necessary thus damaging the prospect of genuine route formation while trying to isolate malicious nodes.

## 4. OUR APPROACH

Our solution eliminates the drawbacks of the naïve solution while also eliminating the malicious effects caused due to flooding.

The core point of the solution explained in Section 3 is that the responsibility of containing the RREQ flooding is shifted to the neighbors of the malicious node(s). Malicious nodes can easily disable RREQ_RATELIMIT and send out as many RREQ packets as possible. Not much can be done to stop the malicious node from doing this. However, the neighbors of this malicious node can work to control the number of fake RREQ packets that are sent, thus preventing the flood from crossing further hops.

To ensure that the flood of fake RREQ packets be controlled and genuine RREQ packets processed with fairness, it becomes necessary for each node to divide its RREQ_RATELIMIT equally among all its neighbors. This bandwidth allotted to each neighbor is updated at regular intervals, depending on the number of *Active Neighbors* that exist in that interval. In any given interval, an *Active Neighbor* is one that has forwarded at least 1 RREQ in the previous 2 intervals. This is done to avoid wastage of bandwidth by allocating it to non existent neighbors.

If $R$ is the RREQ_RATELIMIT capacity of a node, and the number of *Active Neighbors* for the current interval is $N$ then we can define

$$avg_i = (k * R) / N \qquad (1)$$

Here, '$avg_i$' defines the upper limit of the number of RREQ packets that the node can accept from any of its neighbors in that interval. RREQ packets sent by any neighbor beyond $avg_i$ are simply dropped. We introduce '$k$' as the scaling factor to allow an overlap of available bandwidth. It is based on the assumption that not every neighbor will use all the allotted bandwidth in an interval.

As we just mentioned, it is possible that the resources allotted by a node to its neighbors may remain unutilized. Therefore to avoid wastage of bandwidth, our algorithm allows occasional bursts such that some neighbor can utilize almost the entire bandwidth if available. However, if a neighbor violates this burst limit, it is blacklisted for *BT* as explained in the previous section. We denote this burst limit as *peak* and define it as follows

$$peak = α * R \qquad (2)$$

Where $α$ is the maximum tolerable utilization of R by a single neighbor.

(1) and (2) are used together to isolate malicious nodes and control RREQ flooding. We can summarize our approach in the following algorithm

# 5. SIMULATION RESULTS

## 5.1 SIMULATION SETUP

We have chosen a large rectangular area for topology as it enables the formation of long distance routes whose efficiency can be compared between AODV and our modified protocol which we refer to as *ACRR (AODV with Controlled Route Requests).* Parameters 2 and 3 imply a moderate density in the network with approximately 8 to 16 nodes in the interference range. We use NS2 simulator [10] developed by the VINT project with the CMU wireless extensions [11]. The AODV protocol developed by Uppsala University [12] which conforms to RFC 3561 has been used as the base protocol. Random waypoint mobility is assumed in the network. All the results are averaged over 10 simulation runs. The initial placement of nodes each time was random. The vertical bars displayed in the graphs represent the mean with a 95% confidence.

## 5.2 SUCCESSFUL ROUTE FORMATION

This section compares the number of routes formed in AODV and ACRR with respect to 3 parameters-- Malicious Node Count, Network Load and Mobility. The number of routes formed gives an indication of the effectiveness of the routing protocol and the impact of malicious activity.

Figure 2 compares route formation in AODV and ACRR when the malicious nodes are increased from 0 to 6.25%. It can be clearly seen that ACRR performs much better as malicious activity increases.

In Figure 3 we vary the network load and compare the number of routes formed in AODV and ACRR. As network load increases the nodes saturate faster and drop more packets. However, Figure 3 shows that ACRR performs much better than AODV.

Figure 4 shows number of routes formed when mobility is increased from 5 to 20 m/s. With increase in mobility, the routing protocol gets stressed because of increased demand for route requests, specifically route repairs. A quicker route formation will help the connection to be re-established soon. As seen earlier, malicious activity in the network causes degradation in the route forming efficiency. Controlling the flood of RREQ packets will assist in better route formation. Figure 4 shows that ACRR performs better than AODV as the mobility increases.

## 5.3 AVERAGE ROUTE LENGTH

In Figure 5, we compare the route length denoted by the number of hops between source and destination as malicious activity is increased. When malicious RREQ packets spread through the network the genuine packets are denied service and that leads to a slower response. As a result many routes simply time out before they are even created. This is controlled in ACRR and we can see in Figure 5 that ACRR facilitates the formation of a higher number of routes as compared to AODV.

## 5.4 ROUTING OVERHEAD

Routing overhead is the ratio of the amount of TCP data sent to the routing data sent in the network. Since AODV is a reactive protocol, it uses RREQ flooding for route creation. This flooding is justified by the fact that the amount of data sent will be ultimately more and the route formed will be useful for a longer duration of time. Here, we test how well this holds in the presence of malicious data. Figure 6 clearly shows that routing overhead is much higher in AODV as compared to ACRR. When malicious activity increases, the routing packets generated by the malicious node cause network resources to be wasted. Between 2.5% and 6.25% malicious activity the number of RREQ packets sent in AODV is so high that the ratio tends to 0 as shown in the graph, while ACRR successfully manages to have a higher ratio and thus lower overhead. Figure 6 also shows the impact of malicious activity in the network. Lower values of the sent TCP to sent AODV packet ratio imply that less amount of actual data is being sent. Hence in this case we are wasting limited battery power for transmitting useless packets most of the time.

## 6. EXTENSIONS:

All simulations presented above have been done by modifying the AODV protocol. But our algorithm can be applied to any reactive routing protocol for Ad-hoc networks since the underlying method of route formation remains the same.

In our simulations we have assumed homogeneity of the type of mobile devices being used and hence used the same RREQ_RATELIMIT for all nodes. But ACRR can also be extended to work with a heterogeneous network.

## 7. CONCLUSION

We have shown that controlling the flood of route requests in the network by shifting the responsibility of malicious node detection to neighbors helps in improving the overall performance of the network. The RREQ flow control achieved using Eqs. (1) and (2) is much better and flexible than the control achieved when using fixed limits, as done in earlier work.

The extended simulation results presented in this paper further reinforce our result. Our distributed approach does not rely on malicious node information broadcasting and joint decision making, which makes the scheme well suited for Ad-hoc networks.

ACKNOWLEDGEMENT

We would like to take this opportunity to thank Punit Rathod without whom this paper would have been impossible. His innovative ideas and insightful suggestions have been a major contribution toward the successful completion of this paper.

```
Begin
if RREQ is received
    Increment rreq_count for that neighbor
    Calculate the peak and avg values at that instant
    if rreq_count exceeds value of peak
        Blacklist neighbor
        Return
    if rreq_count exceeds value of avg
        Ignore all RREQs till current interval ends
        Return
End
```

**Figure 1: Algorithm for detecting and isolating malicious nodes in the network**

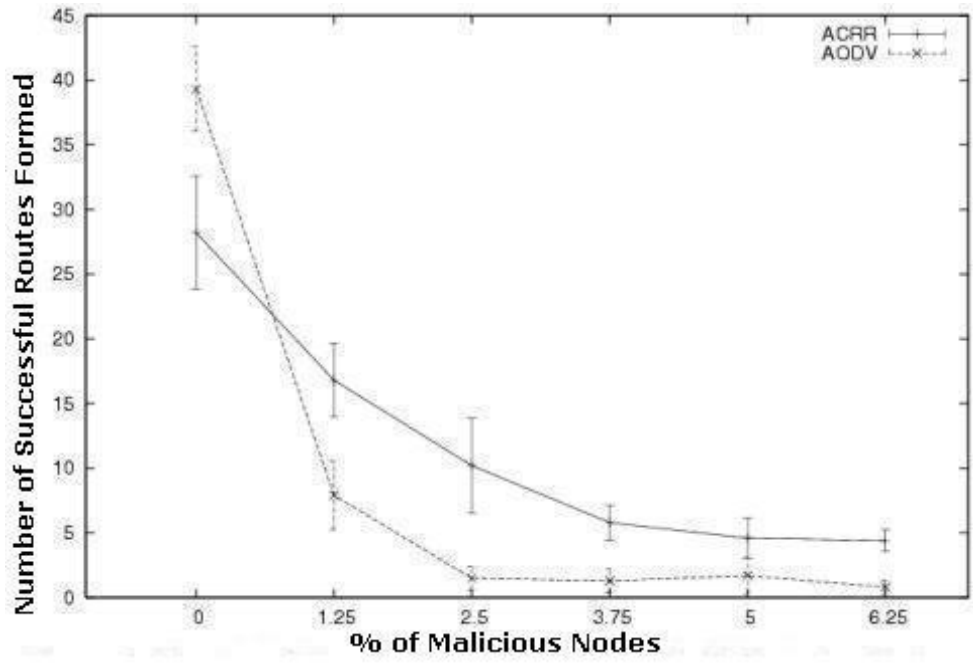

**Figure 2: No. of Successful Routes Formed v/s % of Malicious Nodes**

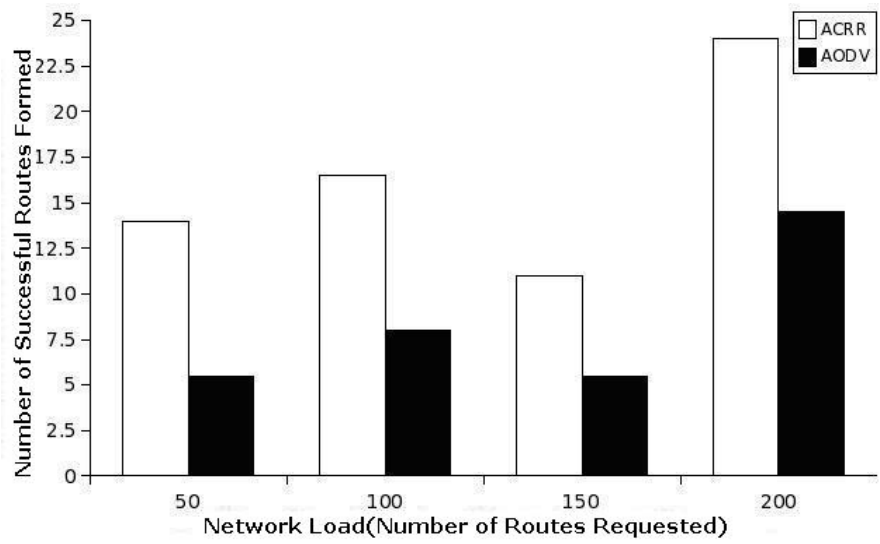

**Figure 3: Number of Successful Routes Formed v/s Network Load. Mobility:10m/s, 2.5%Mal Nodes**

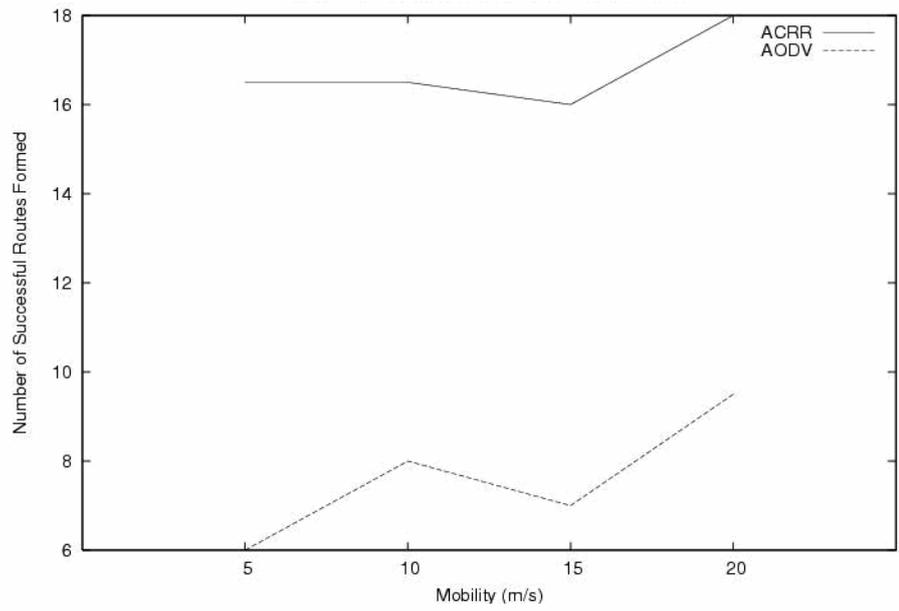

**Figure 4: Number of Successful Routes formed v/s Mobility. Network Load: 100, 2.5% Mal Nodes**

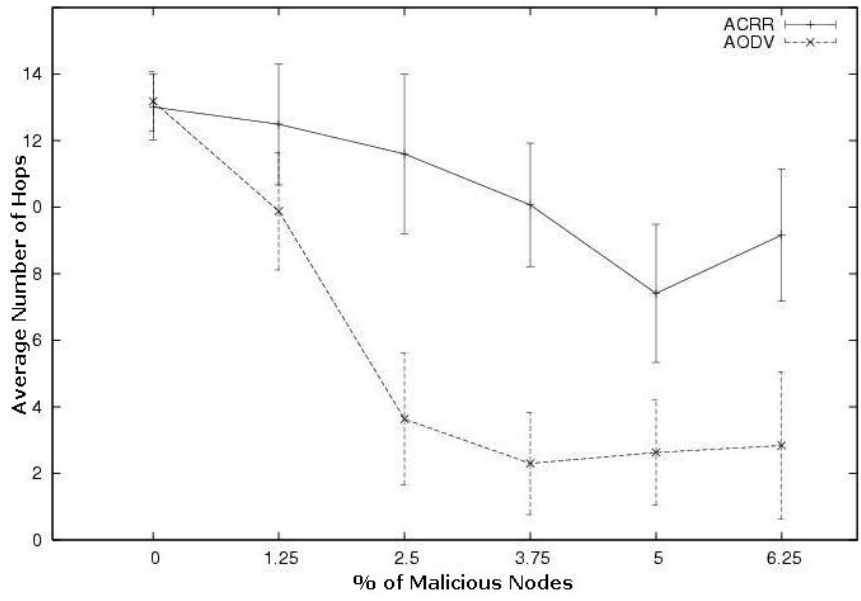

**Figure 5: No. of Hops v/s % of Malicious Nodes**

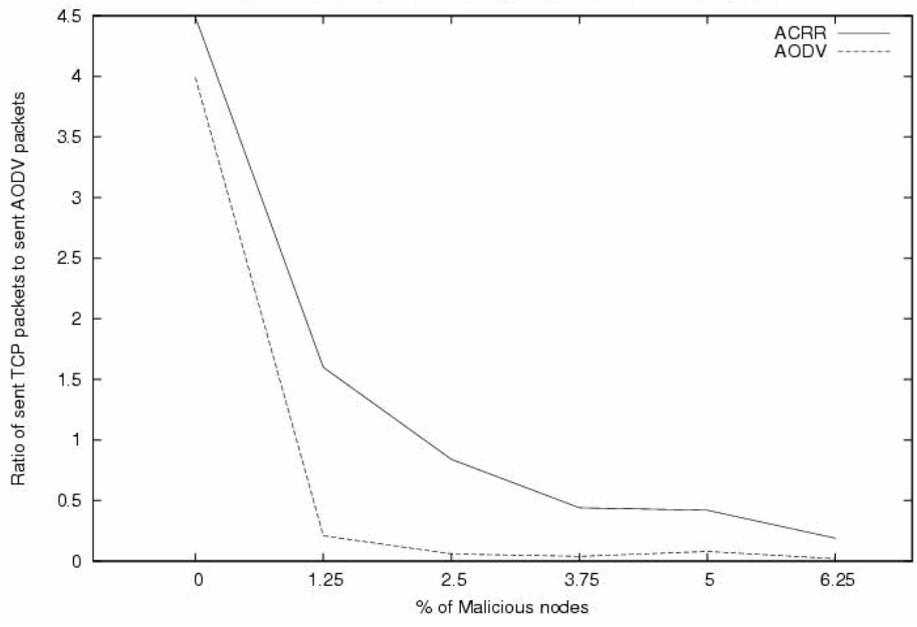

**Figure 6: Routing Overhead v/s % of Malicious Nodes**

| Sr.# | Parameter Type | Value |
|------|----------------|-------|
| 1 | Topology | 5000m$^2$ |
| 2 | No. of Mobile Nodes | 450 |
| 3 | Radio Transmission Range | 250 m |
| 4 | Simulation Time | 50 seconds |
| 5 | Communication Type | FTP |
| 6 | Data Packet Size | 1000 bytes |
| 7 | Confidence Interval | 95% |

**Table 1: Simulation Parameters**